**Pinpointing Lattice-Matched Conditions for Wurtzite Sc$_x$Al$_{1-x}$N/GaN Heterostructures with X-Ray Reciprocal Space Analysis**

Rajendra Kumar[1,2], Govardan Gopakumar[1,2], Zain Ul Abdin[1,2], Michael J. Manfra[1,2,3,4], and Oana Malis[1,2*]

[1]Department of Physics and Astronomy, Purdue University, West Lafayette, Indiana 47907, USA

[2]Birck Nanotechnology Center, West Lafayette, Indiana 47907, USA

[3]School of Materials Engineering, Purdue University, West Lafayette, Indiana 47907, USA

[4]School of Electrical and Computer Engineering, Purdue University, West Lafayette, Indiana 47907, USA

**Abstract:**

Using comprehensive x-ray reciprocal space mapping, we establish the precise lattice-matching composition for wurtzite Sc$_x$Al$_{1-x}$N layers on (0001) GaN to be x = 0.14$\pm$0.01. 100-nm thick Sc$_x$Al$_{1-x}$N films (x = 0.09 - 0.19) were grown in small composition increments on c-plane GaN templates by plasma-assisted molecular beam epitaxy. The alloy composition was estimated from the fit of the (0002) x-ray peak positions assuming the c-lattice parameter of ScAlN films coherently-strained on GaN increases linearly with Sc-content determined independently by Rutherford Backscattering Spectrometry.[29] Reciprocal space maps obtained from high-resolution x-ray diffraction measurements of the $(10\bar{1}5)$ reflection reveal that Sc$_x$Al$_{1-x}$N films with x = 0.14$\pm$0.01 are coherently strained with the GaN substrate while the other compositions show evidence of relaxation. The in-plane lattice-matching with GaN is further confirmed for a 300-nm thick Sc$_{0.14}$Al$_{0.86}$N layer. The full-width-at-half-maximum of the (0002) reflection rocking curve for this Sc$_{0.14}$Al$_{0.86}$N film is 106 arcseconds and corresponds to the lowest value reported in the literature for wurtzite ScAlN films.

---

[*] Author to whom correspondence should be addressed. Electronic mail: omalis@purdue.edu

The III-nitride semiconductors are recognized as an important class of materials, due to their wide bandgaps and highly tunable alloy properties that have been exploited to demonstrate a multitude of electronic and optoelectronic devices.[1-6] Moreover, due to large conduction band offsets and large longitudinal-optical phonon energy,[7] the III-nitrides are promising for designing infrared devices that utilize transitions between strongly confined electronic states, i.e. intersubband (ISB) transitions, such as quantum well infrared photodetectors (QWIPs)[8,9] and quantum cascade lasers (QCLs).[9,10]

Managing lattice mismatch is critical for optimizing the performance and reliability of nitride devices, as well as for expanding their functionality. Lattice-mismatch induced strain leads to the formation of defects such as misfit dislocations, threading dislocations, and stacking faults,[1] that adversely impact material properties limiting device performance. The large lattice-mismatch intrinsic to traditional III-nitrides (e.g. ~11% in InN/GaN and ~2% for AlN/GaN)[13] severely constrains the range of compositions and thicknesses that can be grown by epitaxy before the structure relaxes by generating macroscopic defects.[14,15] Despite $In_{0.17}Al_{0.83}N$ being lattice-matched to GaN, it often exhibits alloy inhomogeneity that affects its properties.[16-21] ISB devices are especially sensitive to structural defects because they consist of multiple quantum well (MQW) structures. In addition to impeding electron transport and generating non-radiative recombination centers, lattice-mismatch induces interface roughening[22,23] that degrades thickness uniformity, leading to ISB transition energy broadening and associated decrease of oscillator strength.[24,25] Since ISB structures are usually several hundred nanometer thick, even minute residual strain can cause the layers to completely relax via extended defects that preclude device applications.

Wurtzite $Sc_xAl_{1-x}N$ has recently become a material of interest for integration with nitride semiconductors because it can be lattice-matched to c-plane GaN. However, the exact composition at which lattice matching occurs is still a subject of debate. Initially, $Sc_{0.18}Al_{0.82}N$[26-31] was believed to be lattice-matched to GaN, but recently, considerably lower lattice-matching compositions have been reported, i.e. $Sc_{0.09}Al_{0.91}N$[32] and $Sc_{0.11}Al_{0.89}N$.[33] The exact composition needed to produce strain-free layers on c-plane GaN substrates must be determined to facilitate complex device designs.

In addition to lattice matching, ScAlN has displayed other interesting properties, such as high dielectric permittivity[34] and large spontaneous polarization and piezoelectric coefficients.[35] ScAlN films also exhibit ferroelectricity[36,37] that has been used to demonstrate novel acoustic wave resonators[38] and high electron mobility transistors.[39,40] Record low sheet resistances and high electron mobilities have also been observed in AlScN/AlN/GaN heterostructures.[41] Our group has reported exceptionally strong near-infrared ISB transitions in $Sc_xAl_{1-x}N$/GaN MQW structures.[42]

In this study, we establish the exact lattice matching conditions for $Sc_xAl_{1-x}N$ on c-plane GaN by examining in detail the structure of a series of epitaxial layers with different alloy compositions in the range x = 0.09-0.19 using high-resolution x-ray diffraction (HRXRD) techniques. The 100-300 nm thick $Sc_xAl_{1-x}N$ layers were grown on quarters of commercially-available semi-insulating 2 μm-thick c-plane GaN templates on sapphire substrates[4] using plasma-assisted molecular beam epitaxy (MBE). We note that the large lattice mismatch between sapphire and GaN results in relatively high defect densities of the GaN templates (nominal threading dislocation < $5 \times 10^8$ cm$^{-2}$) compared to other semiconductor substrates, and these defects are not eliminated through subsequent epitaxial growth.[11,12] Therefore, it is reasonable to compare the measured ScAlN parameters with those of the GaN templates. The substrates were backside coated with a 1 μm layer of tungsten silicide for better thermal coupling with the MBE heater. The substrates were cleaned with trichloroethylene, acetone, and methanol to remove organic impurities, etched in HCl for 10 minutes to remove any excess metal, and then rinsed with deionized water and dried using $N_2$. The substrates were outgassed into an ultra-high vacuum chamber overnight (>12 hours) at 550 °C prior to growth.

High purity Ga, Al, and Sc (99.999%) metal sources were used for the nitride MBE growth to minimize the negative impact of impurities. Active nitrogen was supplied to the substrate by a Veeco Unibulb radio-frequency (RF) plasma source operating at 250 W with a $N_2$ flow rate of 0.25 sccm. A 150 nm GaN buffer layer is first grown at 720°C under Ga-rich conditions (Ga/N ratio of approximately 1.2) that results in a smooth GaN surface with an atomic force microscope (AFM) measured root mean square (rms) roughness of about 0.4 nm over a 5 × 5 µm$^2$ area. The Sc$_x$Al$_{1-x}$N films were grown at a rate of 2.6 nm/min under the conditions identified as optimal in our previous work.[29] The substrate temperature was 550°C as measured by a pyrometer, while the metal-to-nitrogen ratio was approximately 0.8. These conditions result in chemically homogenous ScAlN alloys as demonstrated by scanning transmission electron microscopy and atom-probe tomography.[29]

The structural properties of Sc$_x$Al$_{1-x}$N films were characterized using AFM and HRXRD. The 100 nm-thick Sc$_x$Al$_{1-x}$N films have smooth surfaces with rms roughness less than 0.5 nm over a scan area of 2 × 2 µm$^2$ that is comparable to the rms of the underlying GaN buffer (see Fig. S1 in the Supplementary Material for AFM images). The Sc$_x$Al$_{1-x}$N films with x = 0.092 and 0.192, though, exhibit slightly higher roughness compared to the films with x = 0.134 - 0.160, likely due to increased lattice mismatch with the GaN substrate. Additionally, cracks were observed on the surface of the 100 nm thick Sc$_{0.09}$Al$_{0.91}$N film (see Fig. S2 in the Supplementary Material for optical microscopy image) while Sc$_x$Al$_{1-x}$N films with x = 0.134 - 0.192 were crack free. The presence of cracks on the surface of the Sc$_{0.09}$Al$_{0.91}$N film is a clear indication that this film is under tensile strain and has relaxed by generation of cracks. This finding is in direct contradiction with a previous report that claims x = 0.09 corresponds to the lattice-matched composition.[32]

HRXRD measurements of the (0002) and ($10\bar{1}5$) reflections were performed using a Panalytical Empyrean High-Resolution diffractometer. The Empyrean diffractometer is equipped with a PIXcel$^{3D}$ detector that is capable of functioning as both a point (0D) and a one-dimensional (1D) detector. Reciprocal space maps were acquired using the 1D configuration to facilitate ultrafast mapping, while rocking curves were obtained using the 0D configuration. Additional symmetric $\omega - 2\theta$ scans of the (0002) reflection were taken with a Panalytical X'Pert$^3$ MRD system. All scans indicate only the presence of the wurtzite ScAlN phase; no evidence was found for rock-salt ScAlN or other intermetallic phases. X-ray rocking curves (XRCs) of the (0002) reflection provide information about defect density, while reciprocal space maps (RSMs) of the $(10\bar{1}5)$ reflection reveal the strain state of the films.

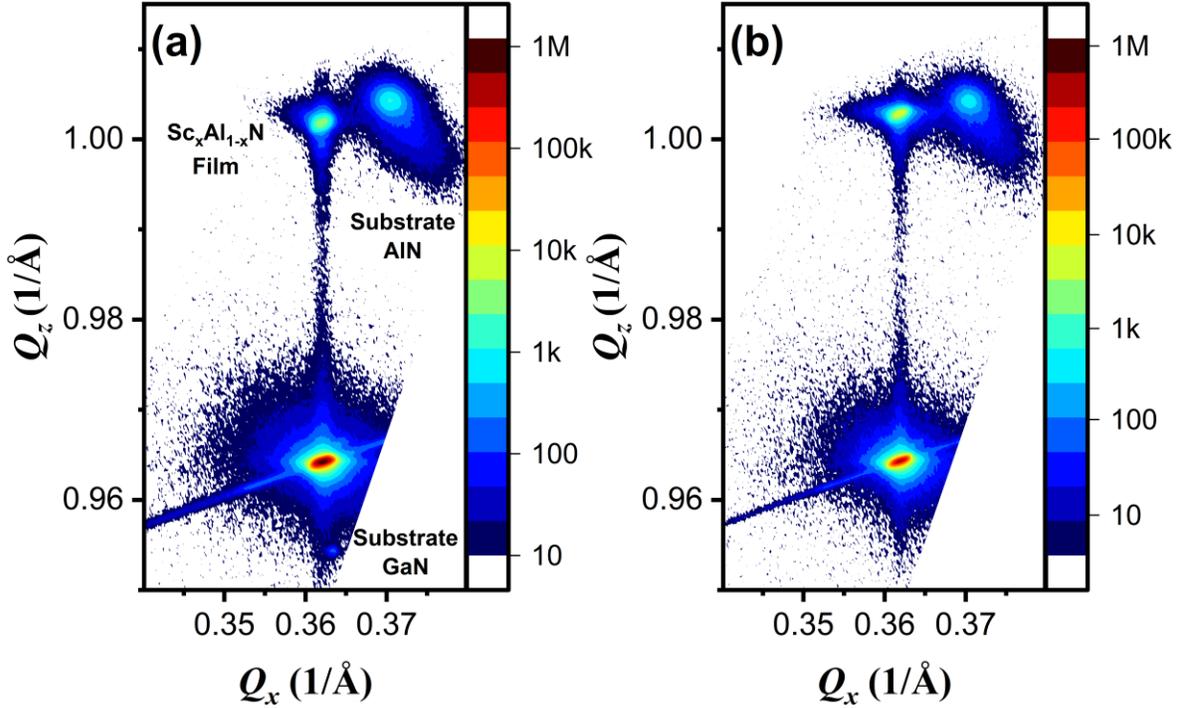

**Figure 1:** Reciprocal space maps of the $(10\bar{1}5)$ reflection of (a) ~100 nm and (b) ~300 nm $Sc_{0.14}Al_{0.86}N$ films. The intensity color scale has units of detected x-ray counts. The maxima for the 2 μm GaN template and a 100 nm AlN nucleation layer originating from the substrate are labeled in (a). The instrumental resolution is reflected in the profile of the $(10\bar{1}5)$ GaN reflection. The diagonal streak observed near the GaN peak is an artifact due to the monochromator.

The out-of-plane $c$-lattice constants derived from RSMs and $\omega - 2\theta$ scans were used to calculate the Sc-molar fraction ($x$) assuming the composition dependence established in our previous work.[29] Chemical analysis by Rutherford Backscattering spectrometry (RBS) performed at Eurofins EAG Laboratories (Sc-composition uncertainty ±1 at.%) of 30-nm $Sc_xAl_{1-x}N$ layers coherently strained on GaN allowed us to determine the following relationship:[29]

$$c = (0.656 \pm 0.027)x + (4.898 \pm 0.004) \text{ Å}.$$

Compositions estimated using this equation are in agreement with expectations from measured Sc beam-equivalent flux used during MBE growth. Furthermore, the alloy composition of a 300 nm $Sc_{0.134}Al_{0.866}N$ layer estimated from HRXRD is in excellent agreement with the independently measured RBS composition of 13.8±1 at.%.

To identify the lattice-matched conditions for ScAlN on c-plane GaN, we examined in detail the RSMs of the $(10\bar{1}5)$ reflection for a series of 100nm thick ScAlN layers in small composition increments between x = 0.09 - 0.19. Figure 1 shows the RSMs for two $Sc_{0.14}Al_{0.86}N$ films of different thicknesses, 100 nm and 300 nm. The full RSMs of all samples are shown in Fig. S3 in the Supplementary Material file. All RSMs show three prominent peaks labeled in Fig. 1(a) that correspond to the 2 μm-thick GaN template, the ScAlN film, and a third AlN peak originating from the substrate. The AlN peak is due to an approximately 100 nm thick layer employed by the substrate manufacturer to reduce dislocation density at the interface between the GaN template and the sapphire substrate. The dominant in-plane and out-of-plane lattice constants for each material can be calculated from the peak reciprocal space coordinates $Q_x$, and $Q_z$, respectively. Coherently strained ScAlN films have the same peak $Q_x$ as the GaN template (i.e. the same in-plane lattice constant). Moreover, the two-dimensional shape of the reciprocal lattice maximum provides additional information about the distribution of interatomic distances. Each reflection is broadened by an instrumental resolution function that is best illustrated by the profile of the GaN reflection. Distortions of the ScAlN peak profile from the instrumental broadening indicate deviation of the interatomic distance distribution from the dominant lattice due to relaxation. Specifically, peak asymmetry can be attributed to partial relaxation due to either tensile or compressive strain.

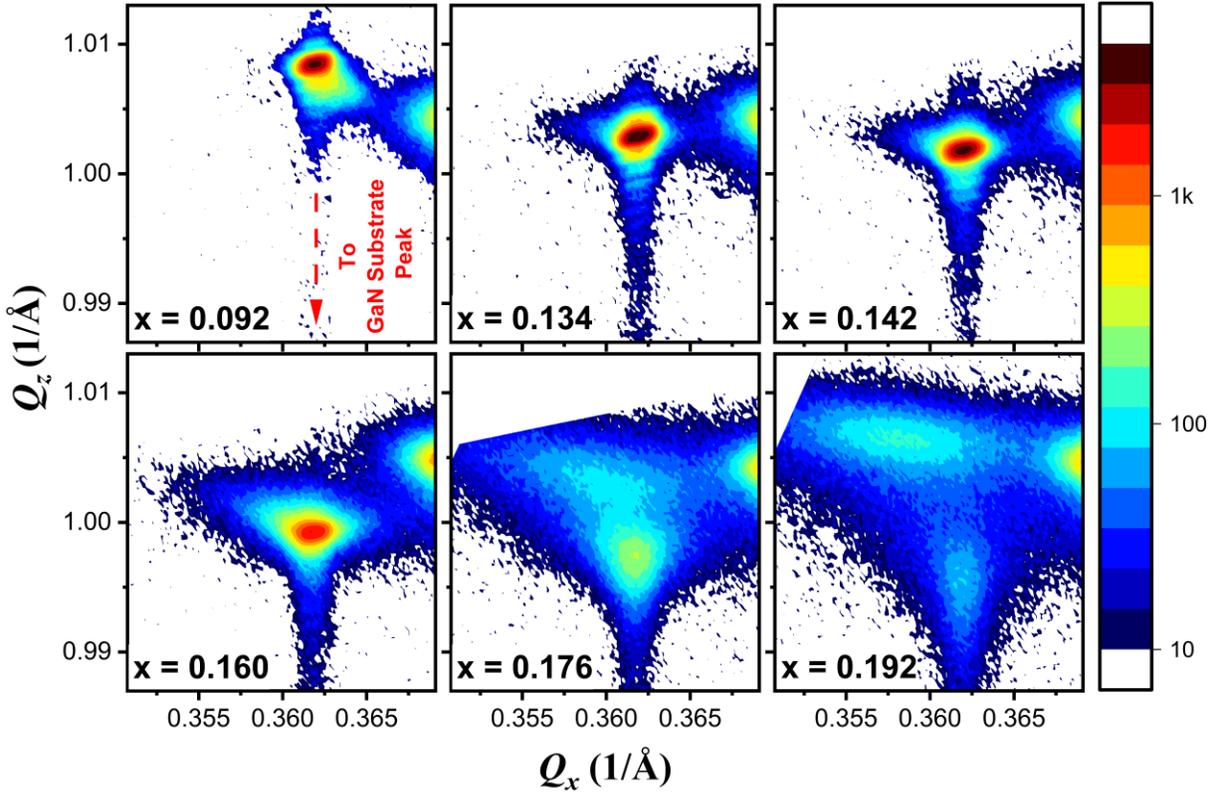

**Figure 2:** Reciprocal space maps of the $(10\bar{1}5)$ reflection for 100 nm $Sc_xAl_{1-x}N$ films of different Sc compositions. The estimated Sc-fraction of each film is shown in the bottom-left corner of each RSM. The intensity color scale has units of detected x-ray counts.

Figure 2 focuses on the $(10\bar{1}5)$ RSMs of the $Sc_xAl_{1-x}N$ layers for six compositions in the probed composition range. The layers with x = 0.134 and 0.142 have the strongest and narrowest reflections and no relaxation-related asymmetry, indicating that lattice matching to GaN is maintained throughout the 100 nm thickness at these compositions. The interference fringes visible along $Q_z$ only for these two compositions indicate the best layer thickness uniformity and can be used to extract the ScAlN layer thickness. In contrast, for the film with x = 0.16, the peak starts to develop an asymmetry towards lower $Q_x$ and higher $Q_z$ that is suggestive of compressive strain relaxation. For even higher Sc-fractions, peak splitting is observed as shown most pronounced for x = 0.192. Peak splitting implies the formation of two distinguishable regions with different average lattice constants within the film. For x > 0.16, the ScAlN layer consists of a region coherently-strained to adjacent GaN buffer, and a relaxed region with larger average in-plane lattice constant and smaller out-of-plane lattice constant likely located near the surface. The x-ray intensities of the strained and relaxed regions within the split peak are a measure of their respective volume fractions. The thickness of the coherently-strained region is, to a first approximation, equal to the relaxation critical thickness.[43] As the Sc composition increases above x = 0.16, the intensity of the relaxed peak increases relative to that of the strained peak. This behavior can only be explained by a reduction of the critical thickness with increasing compressive lattice-mismatch.

Below x = 0.13, the RSM develops a small amount of peak asymmetry towards higher $Q_x$ and lower $Q_z$, corresponding to smaller in-plane interatomic distances, and larger out-of-plane distances, respectively. This is illustrated in Fig. 2 for the $Sc_{0.09}Al_{0.91}N$ film. Additional RSM data is also available in the Supplementary materials file for x = 0.12. The asymmetry of the film peak in RSM along with the formation of cracks indicate relaxation due to tensile strain. Based on analysis of all tested compositions, we therefore conclude that $Sc_xAl_{1-x}N$ with x = 0.14±0.01 is in-plane lattice-matched to c-plane GaN.

To further quantify the differences between ScAlN $(10\bar{1}5)$ RSMs, Figure 3 presents a comparison of the full-width-at-half-maxima (FWHM) of the reciprocal space maxima in directions parallel to the $Q_x$ and $Q_z$ axes. The FWHMs were determined from cross sections passing through the local intensity maxima. For the samples exhibiting peak splitting, both maxima were examined and the values corresponding to the strained and relaxed regions are shown in Fig. 3. The lowest peak widths are achieved for x = 0.092 – 0.142 and are larger, as expected, than the minimum values measured for the underlying GaN template. In this range, the FWHMs increase relatively slowly with Sc composition. We note that even though the minimum FWHM parallel to $Q_x$ is displayed by the film with x = 0.092, the peak asymmetry due to plastic relaxation is not captured in this value because it lies outside the horizontal cross section through the intensity maximum. Films under tensile strain tend to crack rather than undergo plastic relaxation,[43] and as a result, the contribution of the peak asymmetry to the maximum intensity region is less than 10%. The peak widths increase rapidly for x ≥ 0.16. Since the FWHMs are inversely proportional to the correlation length (i.e. coherent x-ray scattering domain), we infer that the density of defects increases rapidly with Sc-composition for x ≥ 0.16, as expected when moving away from lattice-matched conditions.

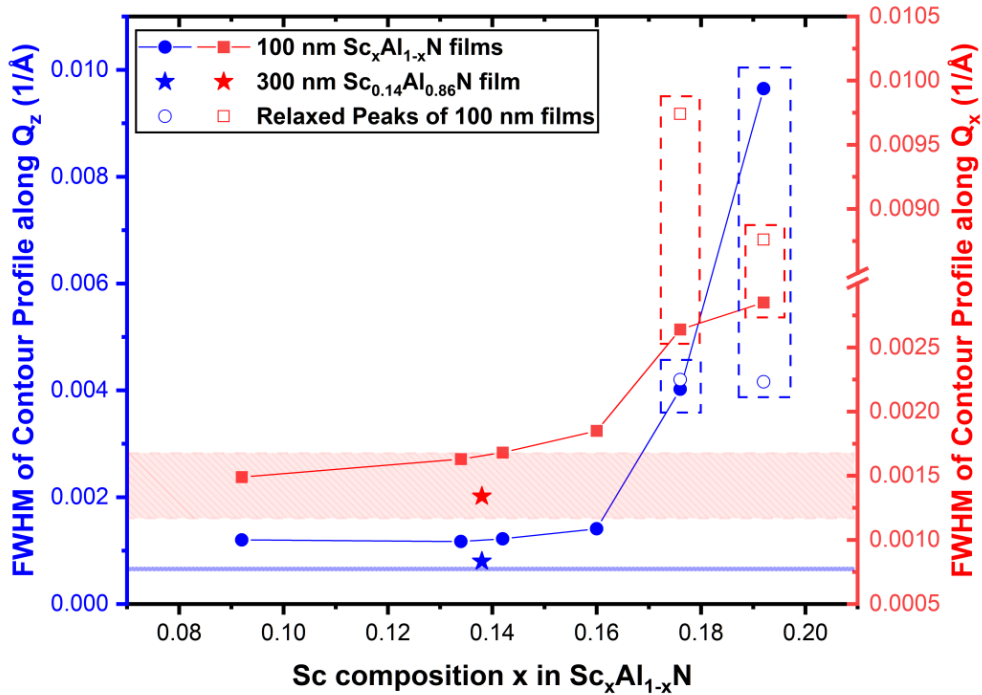

**Figure 3:** Full widths at half maximum (FWHM) values for contour profiles of the film peaks in Figure 2, along $Q_z$ (blue, left y-axis) and $Q_x$ (red, right y-axis). The red shaded region and blue thick line show the range and value, respectively, of the measured FWHM for the underlying GaN substrates. Filled and empty symbols correspond to the strained, and relaxed regions, respectively. Dashed boxes enclose FWHM values for films that exhibit split peaks. The stars are FWHM values for the 300 nm $Sc_{0.14}Al_{0.86}N$ film shown in Figure 1(b).

The lattice-matching condition at x = 0.14 ± 0.01 was additionally tested by growing a 300 nm $Sc_{0.14}Al_{0.86}N$ film to allow further accumulation of stress, if any. The optical micrograph of the film revealed a crack-free surface. Figure 1(b) shows the RSM of the 300 nm $Sc_{0.14}Al_{0.86}N$ film. The single film maximum indicates that layer is strained to the GaN substrate and does not exhibit any plastic relaxation. The FWHMs parallel to the $Q_x$ and $Q_z$ axes are smaller than the widths obtained for ~100 nm films, as shown in Figure 3, suggesting a reduction of defect density relative to the 100 nm films. Using the lattice parameters derived from the RSM peak, the composition was calculated to be 0.134 ± 0.01, whereas independent RBS measurements estimated the composition to be 0.138 ± 0.01. The agreement between these values highlights the reliability of using c-lattice parameters of strained films to determine the composition of $Sc_xAl_{1-x}N$. The AFM rms surface roughness is 1.150 nm over 2 × 2 $\mu m^2$ (see Supplementary materials file) which is significantly higher than the underlying GaN buffer, but somewhat expected given the low growth temperature (550°C) and metal to nitrogen ratio of 0.8.

The overall crystalline quality of the $Sc_xAl_{1-x}N$ films is assessed by measurements of the x-ray rocking curves of the (0002) reflection as shown in Figure 4. The rocking curves were fitted with a Gaussian function to estimate their FWHM values. Measurements of the (0002) XRCs of the underlying GaN substrate yield FWHM values of approximately 55 arcsec, consistent with manufacturer specification of 56 - 73 arcsec. Instrumental broadening, estimated at 20 arcsec, was determined by measuring the rocking curve of the (0006) reflection of sapphire. Remarkably narrow XRC were observed for all ScAlN films. All FWHMs are below the smallest value reported in literature (225 arcsec).[44] Notably, the 300 nm $Sc_{0.14}Al_{0.86}N$ film demonstrated an exceptionally low FWHM of 106 arcsec (Figure 4(b)). The smallest FWHM values reported in literature are listed in the Supplementary Material, to highlight the crystalline quality of the samples discussed in this study.

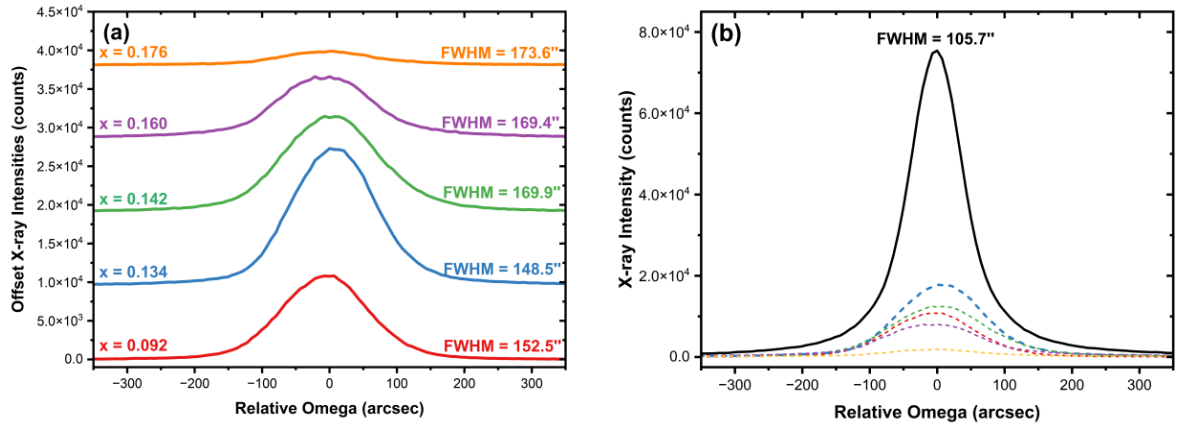

**Figure 4:** XRD rocking curves of the (0002) reflection for (a) ~100 nm $Sc_xAl_{1-x}N$ films and (b) ~300 nm $Sc_{0.14}Al_{0.86}N$ film. The dashed lines in (b) are the rocking curves of the 100 nm films, shown as a reference. The composition (x) and FWHM values are given next to each curve in (a).

In summary, we have systematically explored the lattice matching conditions for wurtzite $Sc_xAl_{1-x}N$ on c-plane GaN in the range x = 0.09 – 0.19. The chemical alloy composition was determined from HRXRD measurements calibrated by RBS and confirmed by additional independent measurements. Our findings indicate Sc composition x = 0.14 ± 0.01 is lattice-matched to GaN. Even a small deviation from this composition (±0.02) leads to peak asymmetry in reciprocal space maps. Further deviation from this composition in the compressive regime (x > 0.16) results in stress relaxation through plastic deformation as evidenced by x-ray peak-splitting. Tensile stress is mainly relaxed through cracks (x < 0.13). Additionally, the MBE-grown ScAlN layers show the lowest (0002) rocking curve widths reported in the literature to date. Specifically, the record FWHM of 106 arcsecond exhibited by a 300 nm thick $Sc_{0.14}Al_{0.86}N$ layer indicates that excellent crystalline quality is achievable under our MBE growth conditions. These results will enable design of exact lattice-matched $Sc_xAl_{1-x}N$/GaN heterostructures and superlattices of arbitrary thickness facilitating optimal performance of novel electronic and optoelectronic devices.

## Supplementary Material

The supplementary material file contains AFM images for ScAlN films discussed in the text, the microscopy image for the 100 nm $Sc_{0.09}Al_{0.81}N$ film, full RSMs for the 100 nm ScAlN films, additional AFM and RSM for a 200 nm $Sc_{0.12}Al_{0.88}N$ layer, and a table of the ScAlN rocking curve widths reported in the literature.

## Acknowledgements

We acknowledge support from the National Science Foundation (NSF). G. G., and O.M. acknowledge partial support from NSF award DMR-2004462.

## Data Availability

The data that supports the findings of this study are available within the article.